\documentclass[useAMS,usenatbib]{mn2e}
\usepackage{psfig, epsf, epsfig}
\title[The Triangulum stream]{Formation of a giant HI bridge
between M31 and M33 from their tidal interaction}
\author[K. Bekki
]{Kenji Bekki${}^1$\thanks{E-mail:
bekki@phys.unsw.edu.au}  \\
       ${}^1$School of Physics, University of New South Wales,
              Sydney 2052, NSW, Australia}

\begin{document}

\date{Accepted, Received 2005 February 20; in original form }

\pagerange{\pageref{firstpage}--\pageref{lastpage}} \pubyear{2005}

\maketitle

\label{firstpage}

\begin{abstract}

Recent observations have discovered a giant HI bridge that appears to connect
between the outer halo regions of M31 and M33. 
We propose that  this HI bridge can be formed as a result
of the past interaction between M31 and M33 based on 
test particle simulations
with different orbits of M31 and M33 for the last $\sim 9$  Gyr.
We show that strong tidal interaction between M31 and M33 about $4-8$ Gyr
ago can strip HI gas from  M33 to form HI streams around M31
which can be observed as a HI bridge if they are projected onto the sky.
We show  that the number fraction of models reproducing
well the observed HI  distribution of the bridge
is only $\sim 0.01$\% 
(i.e., $\sim 10$ among $\sim 10^5$ models) and thus suggest
that the observed structure and kinematics of the HI bridge
can give some  constraints on the past orbits of M31 and M33. 
We  suggest that the observed outer  HI warp in M33
could be fossil evidence for the past M31-M33 interaction.
We also  suggest  that some of high velocity clouds (HVCs) recently found 
in M31 could be HI gas originating from M33.
We briefly discuss other possible scenarios for the formation
of the HI bridge.

\end{abstract}


\section{Introduction}

Recent observational studies on HI gas around M31 based
on the Green Bank Telescope (GBT) have reported
the presence of at least 20 discrete  HI clouds within  50 kpc
of the  M31 disk
(e.g., Thilker et al. 2004).  
Furthermore
possible candidates of HVCs around M31
with heliocentric velocities of $-520$ to $-160$ km s$^{-1}$
have been discovered in a blind HI survey for the disk and halo
regions of M31 (e.g., Westmeier et al. 2007).
Braun \& Thilker (2004, hereafter BT04) have discovered a faint bridge-like HI
structure that appears to connect between  M31 and M33.
Although these recent results have provided  new clues to formation
and evolution of M31 (BT04),
no theoretical works have yet clarified the origin of the HI properties
surrounding  M31, in particular, 
the intriguing bridge-like structure between  M31 and M33.

Corbelli et al. (1989) revealed that the HI gas disk of M33
extends out to twice the optical radius and shows
complicated kinematical properties indicative of 
the presence of a warped disk.
By using a tilted ring model,
Corbelli  \& Schneider (1997) demonstrated that 
the observed distribution and rotation curve profile of HI in the outer
part of M33 are  consistent with the presence of a warped gas disk.
Although they suggested the formation of the gaseous warp in M33
due to tidal force from M31, 
the observed unique HI properties of  M33 have not been discussed
by theoretical and numerical studies in terms of 
the past M31-M33 interaction.

The purpose of this Letter is to first show that
the observed HI bridge between M31 and M33 can be formed
from the past tidal interaction between M31 and M33
based on a larger number ($>10^6$) of
test particle simulations for the past interaction.
We here try to choose  orbital models of M31 and M33
(among a large number of those)
which can reproduce well the observed distribution of the HI
bridge (BT04).  
We therefore do not intend to use hydrodynamical
and chemodynamical simulations that are numerically costly 
and were used in our
previous studies for the formation of HI streams and bridges  around  galaxies 
(e.g., Bekki et al. 2005a, b; 2008).

\section{The model}

The present model is two-fold as follows.
Firstly we derive  the  orbital evolution
of M31 and M33  in the last 9.2 Gyr 
for a given set of the present  three-dimensional (3D) velocities
based on the ``backward integration scheme'' (Bekki \& Chiba 2005; BC05),
in which equation of motion is integrated backward to derive
the past orbits of M31 and M33.
Then we investigate the dynamical evolution
of stellar and gaseous disks of M33 for each of the derived orbits
using {\it test-particle
simulations}.
In this second  step,
we investigate whether the spatial distribution
of  gas particles stripped from M33 due to tidal interaction
between M31, M33, and the Galaxy
can be consistent with the observed HI distribution by BT04
for each of the orbital models derived in the first step.
Throughout this paper,  the terms ``initial'' and ``final''
mean 9.2 Gyr ago (i.e., the time $T=-9.2$ Gyr) and
the present ($T=0$ Gyr), respectively.

\subsection{Orbital evolution}

The total masses of M31 ($M_{\rm M31}$), 
the Galaxy ($M_{\rm G}$), and M33 ($M_{\rm M33}$) are assumed
to be $1.2 \times 10^{12} {\rm M}_{\odot}$,
$1.9 \times 10^{12} {\rm M}_{\odot}$,
and $3.8 \times 10^{10} {\rm M}_{\odot}$, respectively,
which are consistent with observational studies
by Evans \& Wilkinson  (2000),  Wilkinson \& Evans (1999),
and Corbelli \& Schneider (1997), respectively.
The gravitational potential of the Galaxy ${\Phi}_{\rm G}$
is assumed to have the logarithmic potential;
\begin{equation}
{\Phi}_{\rm G}(r)=-{V_0}^2 \ln r ,
\end{equation}
where  $r$ and $V_{0}$ are the distance from the Galactic center
and the constant rotational velocity (= 220 km s$^{-1}$), respectively.
The logarithmic potential with $V_{0}=250$ km s$^{-1}$
is used for M31. These adopted  values for $V_{0}$
 are consistent with observations
(e.g., van den Bergh 2000).
For the adopted mass profile for the above potential,
the mass of a galaxy within $r$ ($M(r)$) can exceed
the adopted total mass of the galaxy at some point.
We thus introduce a  cut-off radius  beyond
which $M(r)$ is constant (e.g., $M(r)=M_{\rm G}$ for the Galaxy).
M33 is represented as a point-mass particle with the mass of $M_{\rm
M33}$. 

\begin{table}
\centering
\begin{minipage}{85mm}
\caption{The ranges of the simulated present  3D velocities of M31 and M33.}
\begin{tabular}{cccc}
Galaxy name & 
{$V_{\rm X}$  
\footnote{Minimum and maximum velocities adopted in orbital models
are shown in units of km s$^{-1}$.}  } &
$V_{\rm Y}$  &
$V_{\rm Z}$   \\
M31 & (-148, 286)  & (-286, 124)  & (-110, 231) \\
M33 & (-165, 214)  & (-206, 168)  & (-126, 174) \\
\end{tabular}
\end{minipage}
\end{table}

We investigate the 3D orbital evolution of M31 and M33 
by adopting  the Galactic Cartesian coordinate system used by BC05
in which the Galactic plane is the same as the $X$-$Y$ plane
and the location of the Sun ($X$, $Y$, $Z$)
is (-8.5, 0, 0) kpc.
The initial distances of M31 and M33 with respect to the center
of the Galaxy are set to be 760 kpc and 795 kpc, respectively
(van den Bergh 2000). Since the radial velocities ($V_{\rm r}$)
of M31 and M33 with respect  to the Galaxy are observed (van den Bergh
2000),
we consider that the following two velocity components
are free parameters determining the 3D orbits of M31 and M33:
(1) $V_{\theta}$, where $\theta$ is the angle between the $Z$-axis and
the position vector (${\bf r}$)
that connects a galaxy (or a particle)  and the Galactic center
and (2)  $V_{\phi}$,  where
$\phi$ is the azimuthal angle measured from $X$-axis to
the projection of ${\bf r}$ 
onto the $X-Y$ plane.

For a given set of $V_{\rm r}$,
we investigate 160000 models with different $V_{\theta}$
and  $V_{\phi}$ for M31 and M33.
Although we have investigated a large number of models ($>10^6$),
we show a set of models in which 
$-200$ km s$^{-1}$ $\le V_{\theta} \le$ $200$ km s$^{-1}$ 
and $-180$ km s$^{-1}$ $\le V_{\theta} \le$ $180$ km s$^{-1}$ 
for M31 and M33, respectively,
and 
$-200$ km s$^{-1}$ $\le V_{\phi} \le$ $200$ km s$^{-1}$ 
and $-180$ km s$^{-1}$ $\le V_{\phi} \le$ $180$ km s$^{-1}$ 
for M31 and M33, respectively.
This set of models is chosen, because a larger fraction of models 
can better reproduce the observed location of the HI gas:
other sets of models with larger $V_{\theta}$ and  $V_{\phi}$ 
can not well reproduce the observations.
It should be noted here that the adopted upper value
of  $V_{\theta}=180$ km s$^{-1}$
is well consistent with the transverse motion suggested
by Brunthaler et al. (2005, B05)  who investigated the proper motion
of M33.
The ranges of the {\it present}
 3D velocities ($V_{\rm X}$, $V_{\rm Y}$,$V_{\rm Z}$,)
for M31 and M33 in this set of models are summarized in the Table 1.

\begin{figure}
\psfig{file=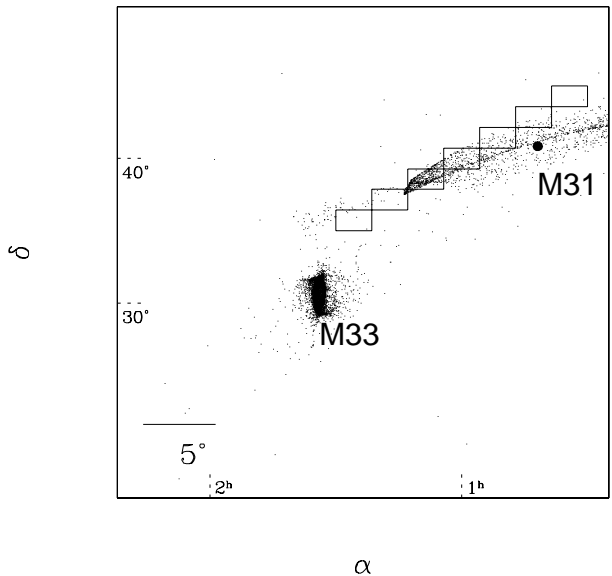,width=8.0cm}
\caption{
The distribution of gas particles stripped from M33 in the
equatorial system (i.e., the $\alpha$-$\delta$ plane)
for the model M1.
The location of M31 is shown by a filled circle for comparison.
The seven rectangular regions shown by solid lines
cover the observed location of the HI bridge (BT04).
These rectangular regions are used for finding the successful
models that can reproduce well the observed distribution of HI gas
around M31.
}
\label{Figure. 1}
\end{figure}

\begin{table*}
\centering
\begin{minipage}{180mm}
\caption{Initial positions and velocities of M31 and M33 
(i.e., 9.2 Gyr ago) adopted for the orbital evolution
in the selected two successful models.}
\begin{tabular}{ccccc}
Model & 
{ ($X$, $Y$,$Z$)  for M31
\footnote{Positions are given in units of kpc.} } &
{ ($V_{\rm X}$, $V_{\rm Y}$,$V_{\rm Z}$)  for M31 
\footnote{Velocities are given in units of km s$^{-1}$.} } &
($X$, $Y$,$Z$) for M33  &
($V_{\rm X}$, $V_{\rm Y}$,$V_{\rm Z}$)  for M33  \\
M1 & (-1682, 1633, 780) & (131, -100, -111) & (-1800, 1617, 677)  
& (180, 85, -21) \\
M2 & (-1046, 398, -1196) & (57, 32, 83) & (-1027, 247, -1196) &
(-24, -102, -29) \\
\end{tabular}
\end{minipage}
\end{table*}

\subsection{Disk evolution}

In test particle simulations,
M33 is assumed to have an exponential disk composed of gas and stars.
Previous observations showed that the size of the HI disk ($r_{\rm g}$)
is at least twice more extended than that of the stellar one ($r_{\rm
s}$)
in M33
(e.g., Corbelli et al. 1989).
We thus consider that (1) $r_{\rm s}=10$ kpc and (2) $r_{\rm g}/r_{\rm
s}>2$. 
For all models,  the observed present inclination angle of $56^{\circ}$ and
the position one  of $23^{\circ}$ (e.g., van den Bergh 2000)
are adopted for initial disk inclinations ($T=-9.2$ Gyr).
The initial rotational velocity of each particle is derived
from $M_{\rm M33}$ and the positions with respect to the center of M33.
Although the final inclination angles of the stellar disk
($T=0$ Gyr) are similar to the initial ones ($T=-9.2$ Gyr),
the final gas disk ($T=0$ Gyr) shows different inclination angles
from those of the initial one owing to the past M31-M33 interaction.
The derived different distributions of
gas and stars, however, are  not inconsistent with observations
by Corbelli \& Schneider (1997) which shows a wap only in
the gaseous component.

For each model,
we investigate whether the final (i.e., T=0, now)
distribution of stripped gas particles
can be similar to the observed one of the HI bridge.
We try to find ``successful models'' in which the following
three conditions are satisfied: (1) at most 50 \% of initial gas
can be stripped to from HI streams, (2) the distribution of stripped
particles can be similar to the observed HI bridge on the
$(\alpha,\delta)$ plane, where $\alpha$ and $\delta$ are right ascension
and declination, respectively,
and (3) the stellar disk of M33 can not be stripped at all.
The above conditions (1) and (2) imply that the tidal radius ($r_{\rm
t}$)
of M33 with respect to M31 should be $r_{\rm s}\le r_{\rm t} \le
r_{\rm g}$ throughout the orbital evolution of M33.
The condition (3) is essentially similar to that adopted by Loeb et al.
(2005) for constraining orbital evolution of M31 and M33.

\begin{figure}
\psfig{file=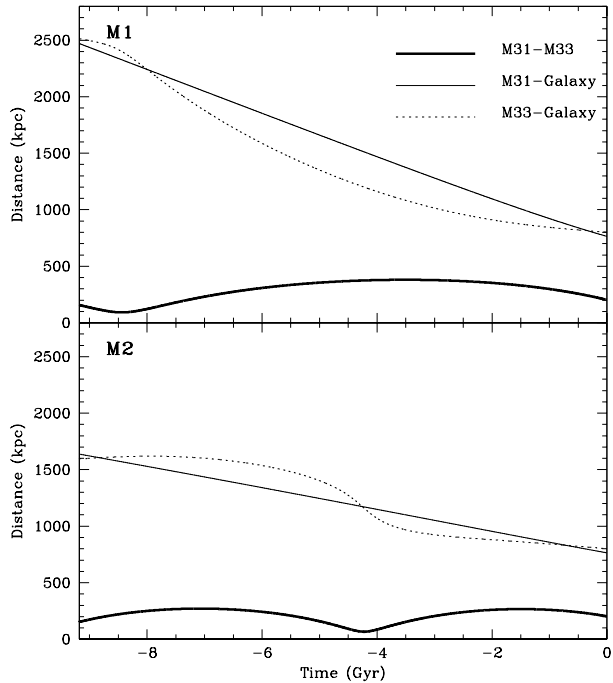,width=8.0cm}
\caption{
Time evolution of distances between M31-M33 (thick solid),
M31-Galaxy (thin solid), and M33-Galaxy (dotted)
for the last $\sim 9$ Gyr in the model M1 (upper)
and M2 (lower). Here the time $T=-9.2$ Gyr and $T=0$ Gyr
represent initial and  final (i.e., now) state
of the model, respectively. 
}
\label{Figure. 2}
\end{figure}

In order to investigate whether the above condition (2)
is  satisfied in a quantitatively manner,
we adopt the following procedure.
We firstly  designate rectangular regions  with the number of $N_{\rm r}$ 
on the ($\alpha,\delta$) plane  so that the rectangular regions can cover
the observed location of the HI bridge (BT04).
Fig. 1 shows the 7 rectangular regions used for investigating
the successful models in the present study. 
We then investigate whether gaseous particle
can be within rectangular regions.
If  a region includes at least one gaseous particle, then
$f_{\rm f,\it i}=1$ ($i=1,N_{\rm r}$)  is allocated: 
otherwise $f_{\rm f,\it i}=0$.
We estimate the following quantity $F_{\rm f}$:
\begin{equation}
F_{\rm f} =\frac{1}{N_{\rm r}} \sum_{i=1}^{N_{\rm r}} f_{\rm f,\it i}
\end{equation}
If $F_{\rm f}$ exceeds  a threshold value of $F_{\rm f, th}$ in a model,
we regard the model as satisfying the above condition (2).

We set $F_{\rm f, th}$ to be 0.7 and thereby derive 
the present 3D velocities of M31 and M33 for the successful
models. This is because we consider that the successful models
explain both the HI bridge and the gas in the northern part
of the M31 halo (Braun \& Thilker 2004): most rectangular
regions need to contain particles in the successful models.
We show the results of the models with
$r_{\rm g}/r_{\rm s}=3.5$ in the present study:
models with smaller $r_{\rm g}/r_{\rm s}$ (e.g., 2.0) do not
show the formation of the HI bridge. 
It is found that only 12 models among 160000 can
satisfy all of the above three conditions for $F_{\rm f, th}=0.7$:
the number of successful models is 293 for $F_{\rm f, th}=0.4$.

For these successful models, we rerun test particle simulations
with particle numbers of 20000 to investigate the details of
the distributions of stripped  gaseous particles on
the $\alpha$-$\delta$ plane. We choose only two successful models
to describe the results in the present study. 
The initial positions and 
velocities of these two models (M1 and M2) are shown
in the Table 2.
Fig. 1 clearly demonstrates that the successful model M1
can really  reproduce the HI bridge between M31 and M33: the detailed 
explanations for Fig. 1  are given  in \S 3.

The present model is more idealized 
(e.g., point-mass model for M33) so that the selection
of best possible orbital models can be feasible. 
If we adopt a more realistic model for the distribution of 
the dark matter halo of M33, the results of the present simulations
could be changed. For example, if we adopt the radial density profile
predicted from hierarchical clustering scenarios 
of galaxy formation (the ``NFW'' profile
;Navarro, Frenk \& White 1996),
the outer part of the M33 disk is more strongly influenced
by the M31 tidal field owing to the weaker gravity of 
the halo in comparison with the present point-mass M33 model.
Although this more efficient stripping in the model
with the NFW profile would increase the total mass
of the stripped HI gas from M33, the selection process
of the best possible model(s)  would not depend so strongly
on the choice of the halo profile: this is because the orbit
of M33 (in particular, the pericenter distance) with 
respect to M31 is the most important parameter for
tidal stripping of the HI gas.

\section{Results}

Fig. 2 shows that (i) M33 is tightly bounded by M31 for the last $\sim$
9 Gyr, (ii) the pericenter distance ($r_{\rm p}$) between M31 and M33 
at the last strong interaction about 8.4 Gyr ago is 92 kpc,
and (iii) both M31 and M33 are 
now approaching the Galaxy for the model M1.
Since the orbit of M33 with respect to M31 is significantly
elongated with the orbital eccentricity ($e$) of 0.6,
the last interaction means the first one.
The  last tidal interaction about 8.4 Gyr ago
is responsible for tidal stripping of gas
and the resultant formation of a HI bridge in this model M1:
the observed HI bridge could be  be a relic of an ancient tidal interaction
between M31 and M33.

Fig. 3 shows that although a significant fraction ($\sim 42$\%) of
gas can be stripped from M33 owing to the M31-M33 interaction,
the stellar disk of M33 can not be strongly disturbed in
the model M1.  The gas disk of M33, however,  appears to show 
a more elongated morphology  with some signs of disturbance.
The gas disk can be significantly more extended than the stellar
one even after the strong tidal interaction with M31 during the last
9 Gyr. The stripped gas can form a tidal stream around M31, which 
appears to have a weak physical connection with
M33 if it is  projected onto
the $X$-$Y$ plane.
Clearly gas particles in the stream are rotating
around M31 and thus show systematic rotation
with respect to  M31.

\begin{figure}
\psfig{file=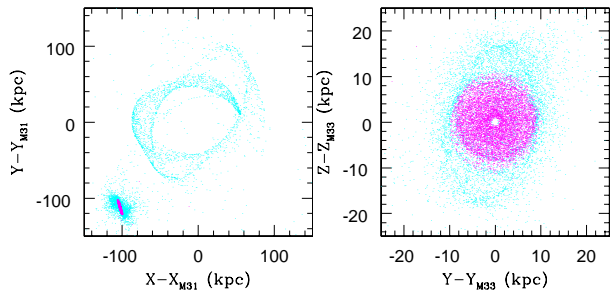,width=8.0cm}
\caption{
Distributions of stellar (magenta) and gaseous (cyan) particles
projected onto the $X$-$Y$ plane (left) and the $Y$-$Z$ one (right)
in the model M1.
The center of frame in the left panel
with a larger field of view  is coincident with
that of M31 so that the tidal stream(s) can be more clearly
seen. The center of frame in the right panel with a smaller field of
view
is coincident with that of M33 so that inner distributions
of gas and stars can be more clearly seen.
}
\label{Figure. 3}
\end{figure}

As shown in Fig. 1, the tidal stream appears to be a giant filament
or a bridge-like structure if it is projected onto the $\alpha$-$\delta$
plane.
The model shows that the stream extends to the north-east  part of
M31, 
which is consistent with the observation (BT04).
The model predicts that there can be no thin tidal streams 
(e.g., leading tidal tails) in the southern part of of M33.
Figs 1 and 3 thus suggest that the observed HI bridge is actually
a tidal stream around M31 viewed edge-on.

Morphological properties of the simulated HI bridges can be different
between the 12 successful models.
Fig. 4 shows the results of the model M2, in which 
M33 can very strongly interact with M31 with small
$r_{\rm p}$ of $67$ kpc  and $e=0.6$ for the first time
around 4.2 Gyr ago (see also Fig. 2 for the orbital evolution).
Owing to the stronger tidal interaction, the final distribution
of gas particles on the $\alpha$-$\delta$ plane appears to be 
more dispersed. However, it is clear that the HI bridge is well
reproduced in this model M2. One of interesting results 
in this model is that some fraction of the stripped gas
appear  to be more uniformly distributed in the surrounding region
of M31.

Thus, these results clearly demonstrate that the observed HI bridge
can be a tidal stream formed from the ancient  tidal interaction
between M31 and M33 possibly about 4-8 Gyr ago. 
The present models,  however, can not provide a robust prediction
on {\it the exact epoch} 
when the HI bridge was formed around M31. Owing to the adopted
test particle simulations {\it without self-gravity of gas},
the observed gaseous warp of M33 can not be formed
in the present models even if M33 strongly interacts with M31.
Given that previous simulations demonstrated the formation
of warps in interacting galaxies (e.g., Tsuchiya 2002),
it is highly likely that 
the past M31-M33 tidal interaction
can also induce the formation of gaseous warps in the outer
part of M33.

We confirm that if the gas disk of M33 is more extended
($r_{\rm g}/r_{\rm s} \sim 3$) {\it when M33 first interacts
with M31},  a bridge-like HI structure can be formed
for a range of orbits of M31 and M33: the M33 gas disk
at the epoch
of the first M31-M33 interaction
could be more extended and more massive 
than the present one. 
The total mass within the bridge depends on the initial
gas mass of M33.
Therefore, the observed column density of the HI bridge
can give some constraints on the initial gas mass of M33,
if the HI bridge is really formed from the M31-M33 interaction.
Column densities and kinematics of HI streams need to be
investigated in fully self-consistent simulations
with gaseous self-gravity,
because gaseous self-gravity can play a key role
in forming local high-density region (e.g., Bekki et al. 2005a).
Our future more sophisticated numerical simulations
with gaseous self-gravity 
will allow us to discuss the observed column density 
and kinematics of the HI bridge in a quantitative manner.

\begin{figure}
\psfig{file=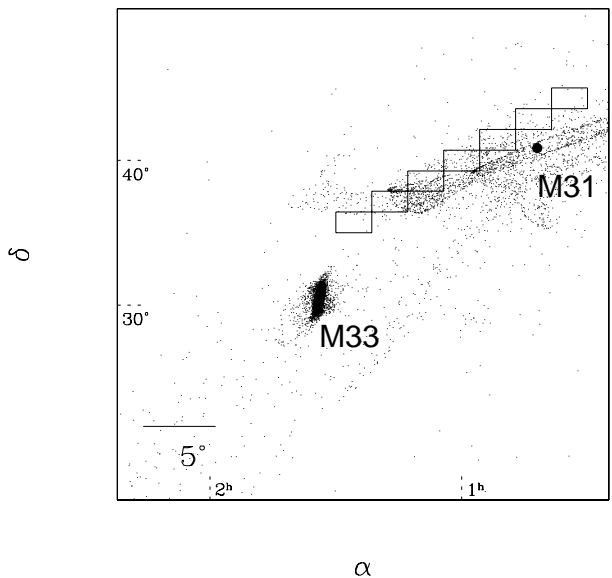,width=8.0cm}
\caption{
The same as Fig. 1 but for the model M2.
}
\label{Figure. 4}
\end{figure}

\section{Discussion and conclusions}

Although the present study has  presented a scenario that 
the observed apparent HI bridge between M31 and M33 is  
a tidal stream of HI gas
around  M31 formed from the past M31-M33 interaction,
this scenario would be only one of several possible scenarios.
For example,
BT04 suggested that the observed bridge can be a   ``cosmic web''
which extends between massive galaxies as predicted in previous
numerical simulations on structure formation.
It would be also possible that the bridge is a tidal stream
originating not from M33 but from other gas-rich dwarfs that may
have already been destroyed by the strong tidal field of M31.
If fully self-consistent hydrodynamical simulations for the
bridge formation in our future studies can also reproduce
the observed sharp HI edge and outer HI warps in M33
(e.g., Corbelli \& Salpeter 1993; Corbelli et al. 1997),
then the present interaction scenario can be regarded as
more realistic and reasonable.

Ibata et al. (2007) have recently found an extended metal-poor
stellar halo around M33 in the deep photometric
survey of M31, though the projected distribution
of the halo is not so clear 
owing to the very limited spatial extent of the survey.
If the stellar halo of M33 has a relatively homogeneous
spatial distribution 
with no signs of disturbance and extends more than 35 kpc from the center
of M33,   then M33 is highly unlikely to have lost
HI gas initially within 35 kpc: the present tidal interaction
scenario
needs to be dramatically modified. 
It would be possible that the observed outer metal-poor halo of M33
can be the very outer part of the M31 stellar halo. 
Future kinematical studies of metal-poor stars
around M33, which can confirm that the stars
belong to M33 rather than to  M31,
will enable us to discuss how far the M33 stellar halo
extends and  thereby assess the viability of the present tidal interaction
scenario.

The present models have shown  that HI gas clouds originating from
M33 can be located in  tidal streams within $\sim 100$  kpc of M31 and
show systematic  rotation with respect to M31.
One of implications from this result is that
at least some of the observed
HVCs around M31 (e.g., Westmeier et al. 2007) can originate 
from M33. 
Metallicities of the M31 HVCs originating from
M33 may well be  as small as those of HI gas in the outer part
of the gas disk in M33. 
Also the HVCs from M33 can have 
systematic rotation with the amplitude
significantly smaller than that of the M31 disk ($V_{\rm c} \sim 250$ km
s$^{-1}$). 
Thus future observational studies on  chemical abundances
and kinematical propertied  of HVCs will help us to reveal
the M31 HVCs originating from M33. 

Although 
the present  gas-rich late-type disk galaxies are observed
to have extended
HI disks (e.g., Broeils \& van Woerden 1994),
it remains observationally unclear  how and when
the extended HI disks were formed. The 
present model M1 (with the first M31-M33
interaction about 8 Gyr ago) would not be reasonable,
if the extended HI disk in M33 was formed relatively recently (4-5 Gyr ago).
It should be thus stressed that the viability  of the present
scenario for the origin of the M31-M33 HI bridge 
would  depend on whether {\it M33 had already developed
its extended HI disk before it experienced the first
tidal interaction with M31}.

Loeb et al. (2005) proposed a proper motion amplitude of
$100\pm 20$ km s$^{-1}$ form M31 based on orbital
models of M31 and M33 and on the observed proper motion
of M33 by B05. 
Recently van der Marel and Guhathakurta (2007) have investigated
line-of-sight velocities of the seventeen M31 satellite galaxies in order to
derive the M31 transverse velocity.
They have found that the Galactocentric tangential
velocity of M31 is highly likely to be less than 56 km s$^{-1}$
and suggested that M31 and M33 is in a tightly bounded system.
These results appear to be consistent with the present tidal
interaction scenario for the HI bridge.
As shown in the present study,
the observed location of the HI bridge
alone can not allow us to provide very precise predictions
on the present 3D velocities of M31 and M33.
We plan to use the observed kinematical data sets for
the HI bridge in order to give much stronger constraints
on the 3D orbits of M31 and M33 by comparing the observations
with more sophisticated numerical simulations.

\section*{Acknowledgments}
I am   grateful to the  referee Rodrigo Ibata
for valuable and constructive comments,
which contribute to improve the present paper.
K.B. acknowledges the financial support of the Australian
Research Council (ARC).

\end{document}